\documentclass[aps,floats,twocolumn]{revtex4}

\usepackage{pifont,colortbl,multirow,graphicx,mathrsfs,makeidx,color,epsfig,
fancyhdr,floatflt,fancybox,calc,amsmath,amsfonts,amssymb,amsthm,latexsym}

\usepackage{epsfig}
\usepackage[latin1]{inputenc}

\newcommand{\nc}{\newcommand}
\nc{\drm}{\rm d} \nc{\Real}{{\rm Re}} \nc{\Le}{{\cal L}}
\nc{\om}{\omega} \nc{\bb}{\begin{equation}} \nc{\ee}{\end{equation}}
\nc{\vecna}{\mbox{\boldmath $\nabla$}} \nc{\ga}{\gamma}
\nc{\ia}{{\bf i}} \nc{\age}{\dagger} \nc{\sig}{\sigma} \nc{\Sig}{\Sigma}
\nc{\var}{\varphi} \nc{\longr}{\longrightarrow} \nc{\al}{\alpha}
\nc{\vare}{\varepsilon} \nc{\C}{C\!\!\!\!C} \nc{\R}{I\!\!R}
\nc{\h}{\hspace*{0.5 cm}} \nc{\hs}{\hspace*}
\nc{\wide}{\widehat} \nc{\ov}{\overline} \nc{\be}{\beta}
\nc{\pa}{\partial} \nc{\cent}{\centerline}
\nc{\vs}{\vspace*} \nc{\Vbf}{\mbox{\boldmath $V$}}
\nc{\Abf}{\mbox{\boldmath $A$}} \nc{\sbf}{\mbox{\boldmath $s$}}
\nc{\jbf}{\mbox{\boldmath $j$}} \nc{\pabf}{\mbox{\boldmath $\partial$}}
\nc{\vbf}{\mbox{\boldmath $v$}} \nc{\abf}{\mbox{\boldmath $a$}}
\nc{\bbf}{\mbox{\boldmath $b$}} \nc{\wbf}{\mbox{\boldmath $w$}}
\nc{\xbf}{\mbox{\boldmath $x$}} \nc{\xibf}{\mbox{\boldmath $\xi$}}
\nc{\Xbf}{\mbox{\boldmath $X$}} \nc{\ubf}{\mbox{\boldmath $u$}}
\nc{\rbf}{\mbox{\boldmath $r$}} \nc{\imp}{\mbox{\boldmath $p$}}
\nc{\sigbf}{\mbox{\boldmath $\sig$}}

\begin{document}

\title{Spin and Madelung fluid\footnote{Work partially supported by INFN--Sezione di Catania}} 

\author{G. Salesi}

\affiliation{Universit\`a Statale di Bergamo, Facolt\`a di
Ingegneria, viale Marconi 5, I-24044 Dalmine (BG), Italy; and
Istituto Nazionale di Fisica Nucleare, Sezione di Milano, via
Celoria 16, I-20133 Milan, Italy}

\email{salesi@unibg.it}

\

\

\begin{abstract}
\noindent Starting from the Pauli current we obtain the decomposition of the 
non-relativistic local velocity in two parts: one parallel and the other 
orthogonal to the momentum. The former is recognized to be the ``classical'' 
part, that is the velocity of the center-of-mass, and the latter the ``quantum'' 
one, that is the velocity of the motion in the center-of-mass frame (namely, 
the internal ``spin motion'' or {\em Zitterbewegung}). Inserting the complete 
expression of the velocity into the kinetic energy term of the classical 
non-relativistic (i.e., Newtonian) Lagrangian, we straightforwardly derive the 
so-called ``quantum potential'' associated to the Madelung fluid. 
In such a way, the quantum mechanical behaviour of particles appears to be strictly 
correlated to the existence of spin and Zitterbewegung.
\end{abstract}

\maketitle

\section{Variational approaches to the Madelung fluid}

 \noindent As is well-known, the Lagrangian for a non-relativistic (NR) scalar particle
  can be assumed to be:
  \bb
  \Le = \frac{i\hbar}{2}(\psi^{\star}\pa_t\psi - (\pa_t\psi^{\star})\psi)
  - \frac{\hbar^2}{2m}{\vecna}\psi^{\star}{\vecna}\psi
  - U\psi^{\star}\psi
  \ee
  where $U$ is the potential due to the external forces, the other symbols
  meaning as usual. \
  Taking the variations with respect to $\psi,\psi^{\star}$, (i.e. working out
  the Euler-Lagrange equations), we get the Schr\"odinger equations
  for $\psi^{\star}$ and $\psi$, respectively.

  The most general scalar wavefunction $\psi \in$  I$\!\!\!$C may be factorized
  as follows:
  \bb
  \psi = \sqrt{\rho}\,{\rm e}^{i\frac{\varphi}{\hbar}}\; ,
  \ee
  where $\rho,\varphi \in$ I$\!$R.
  By this position, eq.(1) becomes:
  \bb
  \Le = -\left[\pa_{t}\varphi + \frac{1}{2m}({\vecna}\varphi)^2
  + \frac{\hbar^2}{8m}\left(\frac{{\vecna}\rho}{\rho}\right)^2 + U\right]\rho.
  \ee
  Taking the variations with respect to $\rho$ and $\varphi$ we obtain$^{[1,2]}$
  the two well-known equations for the so-called Madelung$^{[3]}$
  fluid which, taken together,
  are equivalent to the Schr\"odinger equation, i.e.:
  \bb
  \pa_{t}\varphi + \frac{1}{2m}({\vecna}\varphi)^2
  + \frac{\hbar^2}{4m}\left[\frac{1}{2}\left(\frac{{\vecna}\rho}{\rho}\right)^2
  - \frac{\triangle \rho}{\rho}\right] + U = 0,
  \ee
  where
  \bb
  \frac{\hbar^2}{4m}\left[\frac{1}{2}\left(\frac{{\vecna}\rho}{\rho}\right)^2
  - \frac{\triangle \rho}{\rho}\right] \equiv -\frac{\hbar^2}{2m}\frac{\triangle
  |\psi|}{|\psi|}
  \ee
  is often called ``quantum potential'';
  and
  \bb
  \pa_t \rho + {\vecna}\cdot (\rho {\vecna}\varphi /m)
  = 0.
  \ee
  Eqs.(4),(6) are the {\em Hamilton--Jacobi} and the {\em continuity}
  equations for the probabilistic fluid respectively, and constitute the
  ``hydrodynamical'' formulation of the Schr\"odinger theory.
  Usually, they are not obtained by the above variational
  approach, but by inserting position (2) directly into the Schr\"odinger
  equation and subsequently separating away the real and imaginary parts.
  This second way of proceeding does obey merely to mathematical requirements,
  and does not gives any physical insight of the Madelung fluid.
  On the contrary, our variational approach
  can provide us with a physical interpretation of
  the non-classical terms appearing in Eqs.(3-4).

  The early physical interpretation of quantum potential was forwarded
  by de Broglie's pioneering theory of the {\em pilot wave}$^{[4]}$; in the
  fifties, Bohm$^{[5]}$ revisited and completed de Broglie's approach in
  a systematical way. Sometimes Bohm's theoretical formalism
  is referred to as the ``Bohm formulation of Quantum Mechanics'', alternative
  and complementary to the Heisenberg (matrices and Hilbert spaces),
  Schr\"odinger (wave-functions), and Feynman (path integrals)
  ones. From Bohm's up to present days, several conjectures
  about the origin of that mysterious quantum potential have been put forth,
  by postulating ``subquantal'' forces, the presence of ether, and so on. \ Particularly
  important are the derivations of the Madelung fluid within the
  {\em stochastic mechanics} framework. In such theories, the origin of the non-classical
  term (5) appears as substancially {\em kinematical}. In fact
  to the classical {\em drift} (or {\em translational}) velocity
  $\imp /m$, it is added therein
  a non-classical, stochastic {\em diffusion} velocity 
  (either of {\em markovian}$^{[6]}$ or {\em not markovian}$^{[2]}$ type).
  By adopting markovian--brownian assumptions, the Hamilton--Jacobi eq.(4) is
  obtained in the form of a ``generalized''Newton equation $F = ma$;
  the continuity equation comes out instead from the simple sum of the
  ``forward'' and ``backward'' Fokker--Planck equations
  \footnote{``Generalized'' in that the acceleration $a$
  is defined by means of ``forward'' and ``backward'' time derivatives.}.

  In the present paper we shall correlate, at variance with the above theories,
  quantum potential with the spinning nature of the elementary particles
  constituting matter.
  The starting point is the existence of the socalled
  {\em Zitterbewegung} (ZBW)$^{[7-13]}$ expected to enter any
  spinning particle theories. As is well-known, ZBW is nothing but 
  the {\em spin motion}, expected to exist for spinning particles.
  A spinning particle endowed with ZBW appears as an
  extended-like object, so that the non-classical component
  of the global velocity is actually related
  to the ``internal'' motion [i.e. to the motion observed in the center-of-mass
  frame (CMF), which is the one where $\imp = 0$ by definition].
  The existence of an ``internal'' motion is denounced,
  besides by the mere presence of spin, by the
  remarkable fact that, according to the standard Dirac theory, the
  particle momentum $\imp$ is {\em not} parallel to the velocity:
    $\vbf \neq \imp /m$;
   moreover, in the {\em free} case, while  $[\wide{\imp},
  \wide{H}]=0$  so that $\imp$ is a conserved quantity, quantity
  $\vbf$ is {\em not} a constant of the motion: $[\wide{\vbf}, \wide{H}]\neq
  0 \ (\wide{\vbf} \equiv {\vec {\al}}$ being the usual vector matrix of Dirac
  theory).
Consequently, a decomposition for the global motion, quite analogous to the
one above seen,
in classical {\em plus} non-classical terms, comes out in two famous relativistic
quantum--mechanical procedures: namely, in the
{\em Gordon decomposition}$^{[14]}$ of the Dirac current, and in the {\em
decompositions of the Dirac velocity and Dirac position operators}
proposed by Schr\"odinger in his pioneering works.$^{[7]}$
As we are working in a NR framework, let us recall that
in the literature about ZBW$^{[9-13]}$ it is recognized that the above
decomposition for the velocity holds also in the NR limit,
i.e., for small velocities of the CM \ [$\imp\longrightarrow 0$].
\ In such a way, besides spin and the related intrinsic
 magnetic moment, also another ``spin effect", ZBW, does {\em not} vanish in the
NR theory. Therefore also the Schr\"odinger electron,
being endowed with a ZBW motion, does actually show its spinning
nature, and is not a ``scalar'' particle (as often assumed). As a
matter of fact, when constructing atoms (in the usual NR framework), we have
necessarily to introduce ``by hand" the Pauli exclusion
principle which is related, as known, to spin; and in the Schr\"odinger
equation the Planck constant $\hbar$ implicitly denounces the presence of spin.
All that will be further probed in the next section.
For the moment let us explicitly notice that assuming ZBW is
  equivalent$^{[11,12]}$ to splitting the motion
  variables as follows (the dot meaning derivation with respect to time)
  \bb
  \xbf \equiv \xibf + \Xbf \; ; \ \ \ \dot{\xbf} \equiv \vbf = \wbf + \Vbf \ ,
  \ee
  where $\xibf$ and $\wbf \equiv \dot{\xibf}$ describe
  the motion of the CM in the chosen reference
  frame, whilst $\Xbf$ and $\Vbf \equiv \dot{\Xbf}$ describe the
  internal motion referred to the CMF.
  From an electrodynamical point of view, the conserved electric current is
  associated to the
  helical trajectories of the electrical charge (i.e. to $\xbf$), whilst the center of the
  particle coulombian field is associated to the geometrical centers of such
  trajectories (i.e. to $\wbf$).
As a consequence, it is the charge which performs the {\em total motion},
while the CM undergoes the {\em mean motion} only.

  Going back to Lagrangian (3), it is now possible, starting by the above
  assumptions, to attempt an interpretation of the non-classical term
  $\displaystyle\frac{\hbar^2}{8m}\left(\frac{\vecna\rho}{\rho}\right)^2$ appearing therein.
  Indeed, the first term in the r.h.s. of eq.\,(3) represents, apart from the sign,
  the total energy
  \bb
  \pa_t \varphi = - E \ ;
  \ee
  whereas the second term is recognized to be the kinetic energy $\imp^{2}/2m$ 
  \textit{of} the CM, if one assumes that
  \bb
  \imp = \vecna \varphi.
  \ee
  The third term, that gives origin to quantum potential, may be instead interpreted 
  as the kinetic energy {\em in} the CMF, that is the internal energy due to the 
  ZBW motion, provided that we re-write it in the following form:
  \bb
  \frac{\hbar^2}{8m}\left(\frac{{\vecna}\rho}{\rho}\right)^2 \longr
  \frac{1}{2}m\Vbf ^2,
  \ee
  \bb
  \Le \longr -\left[\pa_{t}\varphi + \frac{1}{2m}({\vecna}\varphi)^2
  + \frac{1}{2}m\Vbf^2 + U\right]\rho.
  \ee
  Eq.\,(11) actually implies
  \bb
  |\Vbf | = \frac{\hbar}{2}\frac{|\vecna\rho|}{m\rho}.
  \ee
  At this point it is easily realized that in Lagrangian (11) the sum
  of the two kinetic energy terms, ${\imp^2}/2m$ and
  $\frac{1}{2}m\Vbf^2$, is nothing but {\em a mere application
  of the well-known K\"onig theorem}.

  In the next section we shall show that assumption (12) can be easily obtained
  from the NR analogue of the so-called Gordon decomposition, that is to say
  from the well-known {\em Pauli current}$^{[15]}$,
  together with a constraint derived from the ``hydrodynamics'' of the Dirac
  equation in the NR limit.\\

  \section{Spin and quantum potential}

  \noindent During the last thirty years Hestenes$^{[13]}$ did
  sistematically employ the Clifford algebras language in the description of
  the geometrical, kinematical and hydrodynamical (i.e., {\em field})
  properties of spinning particles, both in relativistic and NR frameworks.
  He applied the Clifford formalism to Dirac
  and to Schr\"odinger--Pauli theories. In the small-velocity
  limit of the Dirac equation, or directly from Pauli's, Hestenes got
  the following decomposition of the particle velocity field:
  \bb
  \vbf = \frac{\imp - e\Abf}{m} + \frac{\vecna \times (\rho\sbf)}{m\rho}
  \ee
  where the light speed $c$ is assumed equal to 1; $\rho$ is the before-seen
  quantity; $e$ is the electric charge;
  $\Abf$ is the external electromagnetic vector potential;
  $\imp$ is the local momentum, $\imp \equiv
  \rho^{-1}\frac{i\hbar}{2m} [(\vecna \psi^{\age})\psi - \psi^{\age} \vecna \psi]
  $; and
  $\sbf$ is the local spin vector, $\sbf \equiv \rho^{-1}\psi^{\age}\wide{\sbf}\psi$,
    \footnote{Hereafter, every quantity is a {\em local}
    or {\em field} quantity: $\vbf \equiv \vbf (\xbf ;t); \imp \equiv \imp
    (\xbf ;t); \sbf \equiv \sbf (\xbf ;t)$; and so on.}
  where  $\wide{\sbf}$ is the spin operator usually represented by the
  Pauli matrices as:
  \bb
  \wide{\sbf} \equiv \frac{\hbar}{2}(\sig_{x}; \; \sig_{y}; \; \sig_{z})\; .
  \ee
 In this way, the internal ZBW velocity reads:
  \bb
  \Vbf \equiv \frac{\vecna\times (\rho\sbf)}{m\rho}\; .
  \ee
  As a particular case, the {\em Schr\"odinger} one arises; namely, when no external
  magnetic field is present ($\Abf = 0$) and the local spin vector has no
  precession, $\sbf$ is constant in time and uniform in space.
  In this case, we can explicitate the previous equation as follows
  \bb
  \Vbf = \frac{\vecna \rho \times \sbf}{m\rho}\; .
  \ee
As said above, we can notice that, even {\em in the Schr\"odinger
  theoretical framework, ZBW does not
  vanish} (except for the unrealistic case of plane waves, i.e.,
  for the $\imp$-eigenfunctions, for which
  not only $\sbf$, but also $\rho$ is constant,  so that
  $\vecna \rho = 0$). \ Notice also that the continuity equation (6)
  $\pa_t \rho + \vecna\cdot (\rho\imp /m) = 0$ can be still re-written
  in the usual form, namely
  $\pa_t \rho + \vecna\cdot (\rho\vbf) = 0$. \ In fact, since $\vecna\times\sbf=0$,
  we have $\vecna\cdot(\rho\Vbf)=\frac{1}{m}\vecna\cdot(\vecna\rho\times\sbf)
  \equiv\frac{1}{m}\vecna\cdot[\vecna\times(\rho\sbf)]$. Therefore $\vecna\cdot(\rho\Vbf)$, 
  being the divergence of a rotor, is identically equal to zero: as a consequence from Eq.\,(13)
  we get $\vecna\cdot (\rho\imp /m) = \vecna\cdot(\rho\vbf)$.

 By the ordinary tensor language, {\em without employing Clifford algebras},
  we will now show that the decomposition (13) is easily obtained 
  from the familiar expression of the so-called Pauli current (that is, from 
  the Gordon decomposition of the Dirac current in the NR limit$^{[15]}$):
  \bb
  \jbf = \frac{i\hbar}{2m}[(\vecna \psi^{\age})\psi - \psi^{\age} \vecna \psi] -
  \frac{e\Abf}{m}\psi^{\age}\psi +
  \frac{1}{m}\vecna \times (\psi^{\age} \wide{\sbf}\psi).
  \ee
  A spinning NR particle can be described through a Pauli 2-component spinor $\Phi$:
  \bb
  \psi \equiv \sqrt{\rho}\> \Phi
  \ee
  where $\Phi$, if we want to have $|\psi| = \rho$, has to obey the normalization
  constraint
  \bb
  \Phi^{\age}\Phi = 1.
  \ee
Exploiting the expressions for $\sbf$ and $\imp$ introduced at the beginning of this
section and inserting the factorization (18) into the above expression (17) gives just the equation:
  \bb
  \jbf \equiv \rho\vbf = \rho\frac{\imp - e\Abf}{m} + \frac{\vecna\times(\rho\sbf)}{m}
  \ee
  which is nothing but equation (13).

  The Schr\"odinger subcase (i.e., as above said, the case with local spin vector
  constant and uniform) corresponds
  to {\em spin eigenstates}, and then we have to require a wave-function factorizable
  into the product of a ``non-spin'' part $\sqrt{\rho}e^{i\varphi}$ (scalar)
  and of a ``spin'' part $\chi$ (Pauli spinor):
  \bb
  \psi \equiv \sqrt{\rho}\, e^{i\frac{\varphi}{\hbar}}\chi,
  \ee
  $\chi$ being {\em constant with regard to time and space}.
  Now we have
$\imp \equiv \vecna\varphi$ [i.e. eq.(9)],
  $\sbf \equiv \chi^{\age}\wide{\sbf}\chi =$ constant, and, as
  seen above, $\Vbf$ will be given by eq.(16).

Another equivalent way of getting out the decomposition given by eq.(20) is
the one we have recently followed
in ref.$^{[16]}$. In that paper we proposed a new NR velocity operator endowed with
ZBW, starting from which we just obtained the velocity field found above.

Let us finally derive, as promised, equation (12).

Because of the following, known mathematical property of the square
  of the vector product between two generic vectors $\abf$ and $\bbf$:
  \bb
  (\abf \times \bbf)^2 = \abf^2\bbf^2 - (\abf\cdot\bbf)^2\; ,
  \ee
  we have
  \bb
  \Vbf^2 = \left(\frac{\vecna\rho\times\sbf}{m\rho}\right)^2 =
  \frac{(\vecna\rho)^2\sbf^2 - (\vecna\rho\cdot\sbf)^2}{(m\rho)^2}.
  \ee
  Let us now insert in equation (23) the NR
  limit of a constraint found by Hestenes in his hydrodynamical
  formulation of the Dirac theory. Being $\be$ the Takabayasi angle$^{[17]}$,
  Hestenes derived from the Dirac equation, by means of Clifford algebras,
  the following relation:
  \bb
  \vecna \cdot (\rho\sbf) = -m\rho \sin \be.
  \ee
  In the NR limit where $\be \cong 0$ (so that only the two positive-energy
  components do not vanish in the standard Dirac bispinor), equation (24) reduces to
  \bb
  \vecna\cdot (\rho\sbf) = 0.
  \ee
  In the Schr\"odinger case $\sbf =$ const., so that $\vecna\cdot\sbf = 0$; then,
  we can write
  \bb
  \vecna\rho\cdot\sbf = 0.
  \ee
This result can also be derived also in the ordinary algebraic (tensorial)
formalism.
In fact the negative energy component, the so-called ``small''
component of the Dirac (bi)spinor $\chi$, may be written as follows$^{15}$:
$$
\chi \, = \, \frac{\hbar^2}{4m^2}\,|\sigbf\cdot\vecna\varphi|\,,
$$
where $\varphi$ is the positive-energy ``large'' component of the Dirac spinor.
Because, as well known, $\chi \sim \varphi/c$, and consequently
$\rho = \varphi^{\age}\varphi + \chi^{\age}\chi \sim \varphi^{\age}\varphi$,
after few manipulations and approximations we can see that from the smallness
of $\chi$ follows the smallness of $\vecna\rho\cdot\sbf$.
   Putting eq.\,(26) into eq.(23), we easily get
  \bb
  \Vbf^2 = \sbf^2\left(\frac{{\vecna}\rho}{m\rho}\right)^2;
  \ee
  then, since $|\sbf| = \hbar/2$,
  we are finally able to deduce just eq.(12)
  $$
  |\Vbf | = \frac{\hbar}{2}\frac{\vecna\rho}{m\rho}.
  $$
Let us remark that, after having inserted eq.(27) into
  Lagrangian (11),
  the Hamilton--Jacobi and the Schr\"odinger equations can be re-written:
  \bb
  \pa_{t}\varphi + \frac{1}{2m}({\vecna}\varphi)^2
  + \frac{\sbf^2}{m}\left[\frac{1}{2}\left(\frac{{\vecna}\rho}{\rho}\right)^2
  - \frac{\triangle \rho}{\rho}\right] + U = 0,
  \ee
  \bb
  -\frac{2\sbf^2}{m}\triangle\psi = E\psi.
  \ee
  Note that the quantity $2|\sbf|$ {\em replaces} $\hbar$, the latter
  quantity appearing no longer; in a
  way we might say that it is more appropriate to write $\hbar = 2|\sbf|$, rather
  than $|\sbf| = \hbar /2$ \ldots !
With regard to these results, let us recall that in a recent work of ours$^{[18]}$,
even the celebrated de Broglie relation $E = \hbar\om$ is
subjected to a ``spinorial'' interpretation, so that it is therein
substancially ``re-written'' as $E = 2|\sbf|\om$.\\

\section{Conclusions}

 \noindent We have first achieved a Gordon-like
  NR decomposition of the local velocity by the ordinary
  tensorial language.
  \ Secondly, we have derived quantum potential, no longer within the traditional stochastic
  framework, but (without the {\em ad hoc} postulates and the {\em a priori}
  assumptions characterizing stochastic quantum mechanics) by
  relating in a natural way the non-classical energy term to ZBW and spin.
Being quantum potential the ``zero point energy'' of our probabilistic fluid
(in that it is a residual energy, not vanishing even when the CM is at rest
and the external fields are absent), it is quite natural to interpret it
as the ZBW kinetic energy.
The quantum indetermination, to which the zero point energy is strictly related,
results therefore to be connected to the existence of ZBW.
  \ All this carries further evidence that the quantum behaviour
  of microsystems may be a {\em direct consequence} of the existence of spin.
  In fact, when $\sbf = 0$ we consequently have a vanishing quantum
  potential in the Hamilton--Jacobi equation, which becomes then totally
  {\em classical} and Newtonian.
  In this way we are induced to conjecture that
  no really elementary {\em quantum} scalar particles exist,
  but that such particles are always constituted by spinning objects endowed
  with ZBW\footnote{As, e.g., pion composed by quarks.};
  and up to present no contrary experimental evidence has been found.

Finally, we want to recall recent theoretical approaches where the phase space
of the system results to be {\em extended} ---with consequent additional
terms in the hamiltonian structure--- in a way very close to the one followed
in the present work. We refer to the so-called ``semiquantal dynamics''$^{[19-21]}$,
treating the quantum fluctuations of (classically) chaotic systems by means of
essentially classical formalisms, as, e.g., the ``gaussian wave-packet
dynamics'' or the
``time-dependent variational principle''. Both in the latter approach and in another
more recent method$^{[21]}$ based on first-order ``quantum corrections'' to the
classical
equations of motion, the ``semiquantal extended potential'' appearing in the
global effective hamiltonian contains, in addition to the usual classical energy
terms, a ``centrifugal'' barrier endowed with an angular momentum $\hbar/2$.
This last energy term ---common to every physical system and
resulting from the minimum uncertainty condition---
does really correspond to the ZBW kinetic
energy which, as we have shown, is at the origin of the quantum potential.
On the other hand the separation, employed in the quoted papers, of the
canonical variables in classical/centroid \, {\em plus} \, quantum/fluctuations
variables is fully analogous to our decomposition in translational \, {\em plus} \, spin
components. \ From this point of view, our approach and our results may be of
some usefulness for the physical interpretation of those
formalisms, which are rigorous but of essentially analytical
character, and in which nothing is said above the {\em spin} origin of, e.g.,
the quantum corrections or of the quantum ``chaos suppression''.
Therefore it is possible to think that, after further
investigations, all the {\em ad hoc} assumptions of those theories ---from the
extension of phase space to
the ``squeezed coherent state'' assumption, and so on--- may be understood
on the grounds of physical requirements related to the spinning nature of quantum
systems.

\vspace{0.5cm}

\noindent {\large{\bf Acknowledgments}}

\vspace*{0.2cm}
    The author is glad to acknowledge stimulating discussions and
    fruitful cooperation with E. Recami and G. Cavalleri; 
    the collaboration of G. Andronico, G.G.N. Angilella, A. Bugini,
    G. Giuffrida, L. Lo Monaco, E.C. de Oliveira,
    M. Pav\v{s}i\v{c}, W.A. Rodrigues, M. Sambataro, P. Saurgnani, R. Turrisi,
    J. Vaz is also acknowledged. Many references have been
    supplied by E. Recami.
    Special thanks are due to M. Borrometi, L. D'Amico, G. Dimartino,
    F. Raciti, C. Dipietro, R. Salesi and R. Sgarlata.


\begin{thebibliography}{99}

\bibitem {1} F. Guerra and L.M. Morato: Phys. Rev. D{\bf 27} (1983) 1774.

\bibitem {2} G. Cavalleri: Lett. Nuovo Cim. {\bf 43} (1985) 285;
\ G. Cavalleri and G. Spavieri: Nuovo Cim. B{\bf 95} (1986) 194;
\ G. Cavalleri and G. Mauri: Phys. Rev. B{\bf 41} (1990) 6751;
\ G. Cavalleri and A. Zecca: Phys. Rev. B{\bf 43} (1991) 3223.

\bibitem {3} E. Madelung: Z. Phys. {\bf 40} (1926) 332.

\bibitem {4} L. de Broglie: C. R. Acad. Sc. (Paris): {\bf 183} (1926) 447; \
{\em Non-Linear Wave Mechanics} (Elsevier; Amsterdam, 1960).

\bibitem {5} D. Bohm: Phys. Rev. {\bf 85} (1952) 166;  {\bf 85} (1952) 180;
\ D. Bohm and J. P. Vigier: Phys. Rev. {\bf 96} (1954) 208.

\bibitem {6} G.C. Ghirardi, C. Omero, A. Rimini and T. Weber:
Rivista del Nuovo Cimento {\bf 1} (1978) 1, and refs. therein;
\ F. Guerra: Phys. Rep. {\bf 77} (1981) 263, and refs. therein.

\bibitem {7} E. Schr\"odinger: Sitzunger. Preuss. Akad. Wiss.
Phys. Math. Kl. {\bf 24} (1930) 418; {\bf 3} (1931) 1.

\bibitem {8} A.H. Compton: Phys. Rev. {\bf 14} (1919) 20, 247, and
refs. therein; \ W.H. Bostick: ``Hydromagnetic model of an
elementary particle", in {\em Gravity Res. Found. Essay Contest} (1958 and
1961); \ P.A.M. Dirac: {\em The
principles of quantum mechanics} (Claredon; Oxford, 1958), $4^{\rm th}$
edition, p. 262; \ J. Maddox: ``Where Zitterbewegung may lead", Nature
{\bf 325} (1987) 306.

\bibitem {9} M. Mathisson: Acta Phys. Pol. {\bf 6} (1937) 163;
\ H. H\"{o}nl and A. Papapetrou: Z. Phys. {\bf 112}
(1939) 512; {\bf 116} (1940) 153;
\ E.P. Wigner: Ann. Phys. {\bf 40} (1939) 149;
\ M.J. Bhabha and H.C. Corben: Proc. Roy. Soc. (London) A{\bf 178} (1941) 273;
\ J. Weyssenhof and A. Raabe: Acta Phys. Pol. {\bf 9}
(1947) 7; \ M.H.L. Pryce: Proc. Royal Soc. (London) A{\bf 195} (1948) 6;
\ K. Huang: Am. J. Phys. {\bf 20}
(1952) 479; \ H. H\"{o}nl: Ergeb. Exacten Naturwiss. {\bf 26} (1952) 29;
\ A. Proca: J. Phys. Radium {\bf 15} (1954) 5;
\ F. Gursey: Nuovo Cimento {\bf 5} (1957) 784; \
T.F. Jordan and M. Mukunda: Phys. Rev. {\bf 132} (1963) 1842;
\ G.N. Fleming: Phys. Rev. B{\bf 139} (1965) 903;
\ B. Liebowitz: Nuovo Cimento A{\bf 63} (1969) 1235; \
A.J. K\'alnay et al.: Phys. Rev. {\bf 158} (1967) 1484; D{\bf 1} (1970) 1092;
D{\bf 3} (1971) 2357; D{\bf 3} (1971) 2977; \
H. Jehle: Phys. Rev. D{\bf 3} (1971) 306;
\ F. Riewe: Lett. Nuovo Cim. {\bf 1} (1971) 807;
\ G.A. Perkins: Found. Phys. {\bf 6} (1976) 237;
\ D. Gutkowski, M. Moles and J.P. Vigier: Nuovo Cim. B{\bf 39} (1977) 193;
\ J.A. Lock: Am. J. Phys. {\bf 47} (1979) 797;
\ M. Pauri: in {\em Lecture Notes in Physics}, vol. 135
(Springer-Verlag; Berlin, 1980), p. 615;
\ W.A. Rodrigues, J. Vaz
and E. Recami: Found. Phys. {\bf 23} (1993) 459.

\bibitem {10} H.C. Corben: Phys. Rev. {\bf 121} (1961) 1833; \ Phys. Rev.
D{\bf 30} (1984) 2683;
\ Am. J. Phys. {\bf 61} (1993) 551; {\bf 45} (1977) 658;
\ {\em Classical and quantum theories of spinning particles}
(Holden-Day; San Francisco, 1968).

\bibitem {11} A.O. Barut and N. Zanghi: Phys. Rev. Lett. {\bf 52}
(1984) 2009; \ A.O.  Barut and A.J. Bracken: Phys. Rev. D{\bf 23}
(1981) 2454; D{\bf 24} (1981) 3333; \ A.O. Barut and I.H. Duru:
Phys. Rev. Lett. {\bf 53} (1984) 2355;
\ A.O. Barut and M. Pav\v{s}i\v{c}: Class. Quantum Grav.: {\bf 4} (1987) L131;
\ Phys. Lett. B{\bf 216} (1989) 297;
\ M. Pav\v{s}i\v{c}: Phys. Lett. B{\bf 205} (1988) 231; B{\bf 221}
(1989) 264; \ Class. Quant. Grav. {\bf 7} (1990) L187.

\bibitem {12} M. Pav\v{s}i\v{c}, E. Recami, W.A. Rodrigues, G.D.
Maccarrone, F. Raciti and G. Salesi: Phys. Lett. B{\bf 318} (1993) 481;
\ W.A. Rodrigues, J. Vaz, E. Recami and G. Salesi: Phys. Lett. B{\bf 318}
(1993) 623;
\ J. Vaz and W. A. Rodrigues: Phys. Lett. B{\bf 319} (1993) 203;
\ E. Recami and G. Salesi: ``Field theory of the electron: spin and
Zitterbewegung'', in
{\em Gravity, Particles and Space-Time} ed. by P.Pronin and G. Sardanashvily
(World Scientific; Singapore, 1996), pp.345-368;
\ G. Cavalleri and G. Salesi: ``$\hbar$ derived from cosmology and origin
of special relativity and QED'', in {\em Proceedings of ``Physical
Interpretations of
Relativity Theory''} (British Society for the Philosophy of Science; London,
9--12 September, 1994); \ G.Salesi and E. Recami: Phys. Lett. A{\bf 190}
(1994) 137; {\bf 195} (1994) E389.

\bibitem {13} D. Hestenes: {\em Space-time algebra} (Gordon \& Breach; New York, 1966);
\ {\em New foundations for classical mechanics} (Kluwer; Dordrecht, 1986);
\ Found. Phys. {\bf 20} (1990) 1213; {\bf 23} (1993) 365;
{\bf 15} (1985) 63; \ Am. J. Phys. {\bf 47} (1979) 399; {\bf 39} (1971) 1028;
{\bf 39} (1971) 1013;
\ J. Math. Phys. {\bf 14} (1973) 893; {\bf 16} (1975) 573;
{\bf 16} (1975) 556; {\bf 8} (1979) 798;
\ D. Hestenes and G. Sobczyk: {\em
Clifford algebra to geometric calculus} (Reidel; Dordrecht, 1984).

\bibitem {14} J.D. Bjorken and S.D. Drell: {\em Relativistic Quantum Mechanics},
p.36 (McGraw--Hill Book Company; U.S.A., 1964).

\bibitem {15} L.D. Landau and E.M. Lif\v{s}its: {\em Fisica Teorica}, vol.IV:
{\em Teoria quantistica relativistica}, p.153 (Editori Riuniti--Edizioni MIR;
Roma, 1978).

\bibitem {16} G. Salesi and E. Recami: ``About the velocity operator for spinning
particles in quantum mechanics'',
in {\em Proceedings of the ``International Conference on the Theory of Electron''}
(Mexico City, 27-29 September, 1995), \ chapter IV {\em Spin}.

\bibitem {17} T. Takabayasi: Nuovo Cim. {\bf 3} (1956) 233; {\bf 7} (1958) 118;
\ T. Takabayasi and J.-P. Vigier: Progr. Theor. Phys. {\bf 18} (1957) 573.

\bibitem {18} G. Salesi and E. Recami: Found. Phys. Lett. {\bf 10} (1997) 533.

\bibitem {19} A.K. Pattanayak and W.C. Schieve: Phys. Rev. Lett. {\bf 72} (1994)
2855; \ Phys. Rev. E{\bf 50} (1994) 3601.

\bibitem {20} F. Cooper, J.F. Dawson, D. Meredith and H. Shepard: Phys. Rev.
Lett. {\bf 72} (1994) 1337.

\bibitem {21} B. Sundaram and P.W. Milonni: Phys. Rev. E{\bf 51} (1995) 1971.

\end{thebibliography}
\end{document}